\documentclass{osa-article}
\usepackage[utf8]{inputenc}

\journal{ao}

\articletype{Research Article}

\usepackage{lineno}
\begin{document}

\title{Increased phase precision of spatial light modulators using irrational slopes: Application to attosecond metrology}
\author{Geoffrey R. Harrison, Tobias Saule, Brandin Davis, Carlos A Trallero-Herrero*}
\date{May 2022}

\address{Department of Physics, University of Connecticut, Storrs, Connecticut 06269, USA}

\email{\authormark{*}carlos.trallero@uconn.edu} 

\begin{abstract}
The ability of spatial light modulators (SLMs) to modify the amplitude and phase of light has proved them invaluable to the optics and photonics community.
In many applications the bit-depth of SLMs is a major limiting factor dictated by the digital processor. As a result, there is usually a compromise between refresh speed and bit-depth.
Here we present a method to increase the effective bit-depth of SLMs which utilizes a linear slope as is commonly applied to deal with the zeroth order effect. 
This technique was tested using two interferometric transient absorption spectroscopy setups. Through the high harmonic generation in gases producing a train of attosecond pulses and harmonics from solids in the ultraviolet, two pulses are generated that interfere in the far field providing a measurement of the optical phase. An increase in the precision far beyond the limit dictated by the digital processor in the bit-depth was found.
% Note: The abstract should be limited to approximately 100 words. It is currently 98 words
\end{abstract}

% \maketitle

\section{Introduction}

% SLMs are cool because...
% They are used for all sorts of things ...
% one of the inherent limitations is bit-depth ...

% generally SLMs are used with a zeroth order offset for obv reasons...
% also for less obv reasons cite 
% Here we present a technique which uses this things everybody does already to allow sub-bit phase effects and increase effective bit-depth

% One thing you can mention in the intro is that the emphasis so far has been in fast SLMs but not on high resolution ones but, for fine optical manipulation such as high momentum or fine delay control, a higher bit depth is desired.

%Intro + What is an SLM
For the past few decades spatial light modulators (SLMs) have found widespread use in the optics and photonics community because of their ability to readily manipulate the properties of light, e.g. amplitude and phase.
They are used extensively in applications as diverse as optical microscopy\cite{Maurer2011,Gould2012}, imaging though scattering media\cite{Aulbach2011,Tzang2019,Sanjeev2019}, material processing\cite{Yu2016,Hasegawa2014}, quantum entanglement experiments\cite{Mair2001,Leach2010,Fickler2012}, 3-D displays\cite{Liu2011,Kozacki2016,Yamaguchi2016}, optical tweezers\cite{Pollari2015,Curtis2002,Padgett2011}, beam shaping\cite{Clark2016,Ngcobo2013}, and even everyday applications like data storage\cite{Hesselink2004,Li2015}.
While there are many advantages to them, most SLMs have common limitations that are inherent to the digital technology used to drive electro-optical changes in the modulator. Thus, besides the number of pixels which determines the spatial optical resolution, the pixel refresh speed and phase resolution are completely determined by the bit rate of the digital infrastructure.
Other known issues present in SLMs such as spatial phase variance and non-linear phase response can be mitigated by time consuming calibration steps in the form of a wavefront correction mask and phase look up table respectively. 
Ultimately, however, the bit-depth and spatial resolution, which are intrinsic properties of the device, limit how precisely and with what spatial resolution a phase can be adjusted.
One application that is significantly impacted by the bit-depth is the temporal control of attosecond pulses \cite{Tross_Interferometer} and its application to interferometric transient absorption spectroscopy (ITAS) as described in \cite{Jan:orbital_spectroscopy}.
ITAS systems offer some of the most precise phase measurements, with sub-attosecond precision, by controlling the optical phase of the fundamental which translates into a time delay/advance of a train of attosecond pulses. However the bit-depth defines the minimum time resolution, i.e. time step size, that can be measured. 
As an example, for an 800 nm laser incident on an 8 bit-depth SLM with a $2\pi$ phase range the minimum temporal step size is $T_{800}/2^8\approx10$~as (as = attosecond = $10^{-18}$s), with $T_{800}$ the period of the 800~nm pulse.
As ultrafast physics moves into the realm of zeptoseconds (zs = $10^{-21}$s) achieving sub-attosecond time steps is clearly desirable \cite{Grundmann2020}. As an example, to sample dynamics with sub 100 zs steps would require an SLM with over 16 bit-depth which is not readily available.

Another limitation of SLMs is the zeroth-order beam \cite{Zhang2009} which co-propagates with the desired light but is unaffected by the SLM.
While some applications can tolerate this issue, others like holographic data storage\cite{Nobukawa2021} and ITAS, need to address and suppress this effect to improve performance.
The simplest and most common way to correct for zeroth order is to superimpose a linear phase slope on top of the desired phase mask, tilting the beam away from its original direction and separating the two beams in the far field\cite{Jesacher2014}. 
This linear slope has just recently been shown to have another benefit, namely a reduction in temporal phase fluctuations\cite{Nobukawa2021} which is predicted to improve the quality of holographic data storage systems.

Here we present a method to mitigate the bit-depth limit and increase the phase precision of SLM devices.
Similar to phase fluctuation suppression\cite{Nobukawa2021} this technique is based on the addition of irrational linear slopes in addition to a desired phase. 
That leads to an increase in the effective bit-depth of the device via an effect similar to volume averaging.
This method is verified using precise ITAS setups where linear slopes, combined with an overall phase offset were used to achieve step sizes in the time delay of attosecond pulses 3-6 times smaller than what would be achievable otherwise. % predicted by the bit-depth of the device.
We believe this method could be used to gain multiple orders of magnitude more precision, beyond what our system could measure.
This technique was tested with SLMs from two major producers, Meadowlark and Hamamatsu, to demonstrate the generality of its use.

%Talk about the testing briefly
% These results were found while using an Interferometric Transient Absorption Spectroscopy (ITAS) setup as described in \cite{Tross_Interferometer}.
% These systems, along with many others which rely on SLMs \textcolor{red}{cite stuff}, are heavily limited by the bit-depth of the SLM, making more expensive high bit-depth SLMs a necessity.
% For ITAS systems, specifically, the bit-depth acts as a minimum delay step size for the probes, limiting the time resolution of these setups. 
% Here we present a technique, based on volume averaging, that allows SLMs to be used with much higher precision than would be expected from their bit-depth.
% With that the delay added to multiple pixels can be averaged together allowing for a net delay of fractions of a bit.

%%%%%%%%%%%%%%%%%%%%%%%%%%%%%%%%%%%%%%%%%%%%%%%%%%%%%%%%%%%
\section{Experimental setup}\label{setup}

% Describe the setup:
% \begin{enumerate}
% \item basics ie laser desc, SLM desc, two beams into gas jet into spectrometer used to measure phase but not necessary for the improved bit-depth
% \item also tested in solids where the experimental setup is ...
% \item describe how the delay is measured from the interference fringes
% \end{enumerate}

\begin{figure}
\centering\includegraphics[width=14cm]{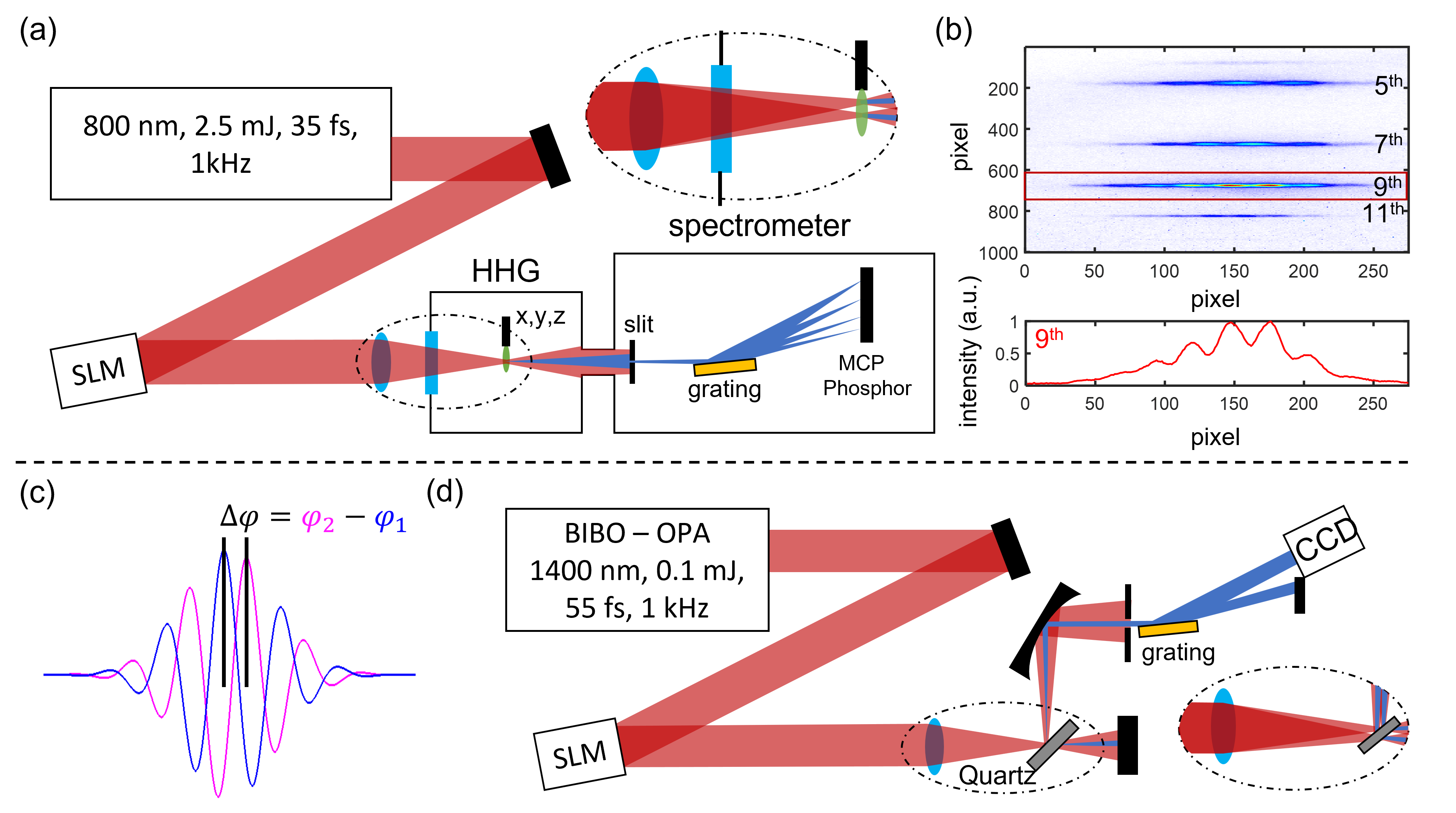}
\caption{
\textbf{(a)} Sketch of the experimental setup using the Meadowlark SLM. 
A SLM is used to separate the incoming beam into two phase-locked pulses.
Both pulses then generate spatially separated but phase-locked harmonics which interfere in the far field and are detected in a spectrometer.
\textbf{(b)} A sample image from the gas spectrometer showing the \(5^{th}\) to \(11^{th}\) harmonics with fringes.
An integrated lineout of the \(9^{th}\) harmonic, outlined in red, is shown below.
\textbf{(c)} A phase offset in one focus leads to a difference between both fundamental foci which in turns creates a phase difference between the harmonics.
\textbf{(d)} Sketch of the experimental setup used for phase measurements in solids using a Hamamatsu SLM. A small portion of one of the signals from an OPA is sent onto the SLM at a steep angle.
The diffracted beam is then focused with a calcium fluoride lens onto the quartz sample where harmonics are generated in the reflection geometry. Harmonics are collimated with a cylindrical mirror and sent onto the camera where spectral fringes are recorded.
} \label{setup_fig}
\end{figure}

To test this technique we measured the phase applied by two SLMs using a self-referencing interferometer for attosecond pulse trains as described in \cite{Tross_Interferometer}.
In short, two intertwined linear slopes create two identical foci after a lens. Each focus generates high harmonics (HHG) in gases \cite{Corkum1993,Chang2016} or lower harmonics in solids \cite{Grynberg2010} employing different setups. 
The experimental setup for the gas experiments is shown in Fig.\ref{setup_fig}(a).
790 nm, 2.5 mJ, 35 fs pulses from our titanium-sapphire laser were separated by a XY Phase Series 512L Meadowlark SLM into two beams that are focused through a 300 mm lens into two foci in an ethylene gas jet.
Since the masks for the two foci already use linear slopes no zeroth order offset was used for these experiments.
The SLMs used in these experiments are anti-reflex coated so the zeroth order is minimal and is not intense enough to generate the harmonics examined here.
% The zeroth order beam creates a third focus but it is not intense enough to 
This SLM has a resolution of 1920x1152 pixels covering an area of 17.6 by 10.6 mm and an 8 bit-depth.
Harmonics are generated at both foci and are sent into a spectrometer where they interfere spectrally (still yielding discrete harmonics) and spatially creating fringes in each harmonic (Fig.\ref{setup_fig}(b)).
The spectrometer uses a laminar, curved, variable spacing extreme ultraviolet (XUV) grating, followed by a micro channel plate phosphor stack and a low noise camera (Hammamatsu Orca Flash).
The phase difference between the two XUV beams, shown in Fig.\ref{setup_fig}(c), can be controlled through the SLM and measured directly from the interference fringes.
A phase offset $\Delta \phi$ given to one of the fundamental beams translates into a phase advance of the harmonic $q$ given by $q\Delta \phi$.
%Some transition could help here
% The phase added by SLMs is often described in units of radians, however, when talking about multiple different SLMs working across multiple different wavelengths the most natural unit to use is bits.
The phase added by SLMs is often described in units of radians, however, when comparing phase masks of different SLMs the most natural unit to use is bits.
Using bits allows us to talk about effects that apply to all SLMs regardless of the central wavelength of design, lookup tables etc.
Conversion to radians from bits and vice versa is done by noting that \(2\pi\) rads of the central wavelength of operation is usually equal to \(2^{\mbox{bit-depth}}\) bits.

Experiments were repeated using a Hamamatsu X10468 SLM which has a resolution of 792x600 pixels covering an area of 15.8 by 12 mm with 8 bit-depth.
Since this SLM operates at a different wavelength, 1400 nm, a separate but similar setup, working with solids, was used to test it (Fig. \ref{setup_fig} (d)).
% This setup is made for similar interferometric experiments but uses harmonics generated in solids instead of gases.
In this setup 790~nm pulses from our laser are sent into an optical parametric amplifier (OPA) where they generate 100~\(\mu\)J, 55~fs 1400~nm pulses. 
Similar to before, these pulses are split by the Hamamatsu SLM and then focused with a calcium fluoride lens onto a quartz sample where harmonics are generated in the reflection geometry\cite{Lu2019}.
Harmonics are spectrally separated by a flat grating and the interference fringes for each harmonic are recorded by a Mightex camera.
%This setup works with lower harmonics for ease of use and then doesn't need to be in vacuum

\indent Both setups use similar masks to split the incoming beam into two foci and then add a phase offset to one of those foci. Because of the unambiguous relation between phase offset and delay of the harmonics we will refer throughout the paper to phase offsets and phase delays indistinguishably.

%%%%%%%%%%%%%%%%%%%%%%%%%%%%%%%%%%%%%%%%%%%%%%%%%%%%%%%%%%%%%%%%
\section{Limits of SLM bit-depth}\label{steps}

% Explain how masks are generated and how a delay is added to one beam.

% \begin{enumerate}
% \item Describe mask slopes + 2pi wrapping
% \item Briefly describe how masks are intertwined for these tests
% \item How a delay is added to each pixel
% \item How bit-depth relates to delay + would expect 1 bit = x rads = y attosecs to be smallest possible delay
% \end{enumerate}

\begin{figure}[ht!]%
\centering
\includegraphics[width=14cm]{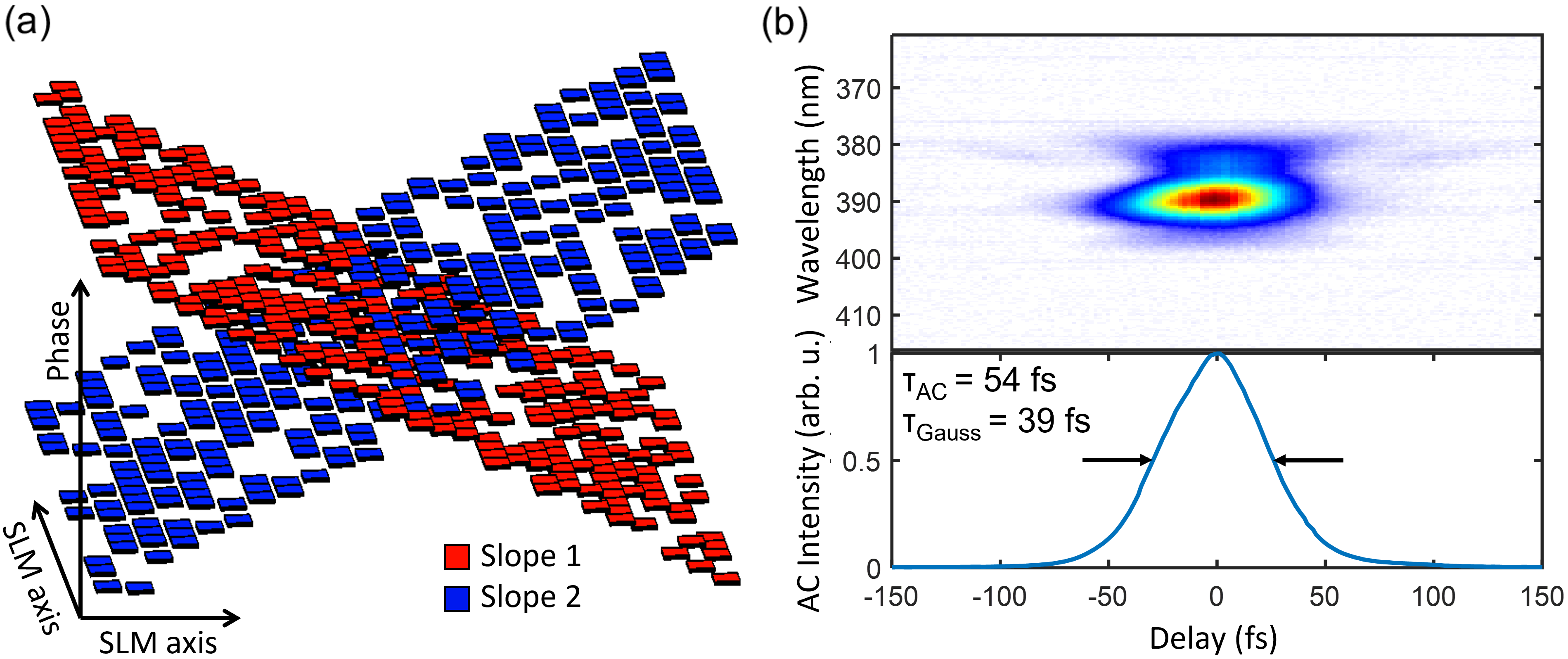}
\caption{
\textbf{(a)} A simplified schematic of how the SLM masks used in these experiments are generated.
The slopes have equal but opposite angles and each has half of the pixels randomly assigned to it.
\textbf{(b)} A FROG trace and autocorrelation of the pulses after being split by the SLM. 
They show an autocorrelation full width at half maximum, $\tau_{AC}$, of 54 fs corresponding to a gaussian pulse width, $\tau_{Gauss}$, of 39 fs. 
% The SLM adds some dispersion and increases the width from 35 fs, but the pulse is still short enough to generate the harmonics required for these experiments.
} \label{new_fig}
\end{figure}

The masks used in these experiments were made of two intertwined linear slopes each applied across the entire SLM. 
The linear slopes split the beam into two angled wavefronts which focus 300 $\mu m$ away from each other after being focused by a lens. 
As can be seen in Fig. \ref{new_fig}(a) the masks are made by randomly assigning each pixel to one of the slopes.
These random masks are used instead of other patterns, such as checkerboards since they have better diffraction efficiency.
As expected, these masks add some dispersion because the SLM applies slightly different slopes to different wavelengths due to the wavelength dependent phase of each pixel. 
However, we find this dispersion to be minimal as quantified by a frequency resolved optical gating (FROG)\cite{Trebino1993} shown in Fig. \ref{new_fig}(b) measured from a beam split by the SLM as described above. 
The pulses are stretched from 35 fs to 39 fs remaining short enough to generate harmonics for ITAS.
Phase delays between the two beams can be created by adding a constant offset to every pixel assigned to one of the slopes.
% The masks used in these experiments were made of two intertwined linear slopes $L_1(x)$, $L_2(x)$, each applied across the entire SLM which create foci 150 \(\mu m\) away from the zeroth order focus in opposite directions.
% Each pixel in the masks is randomly assigned to one of the two slopes to increase diffraction efficiency when compared to checkerboard patterns. 
% Phase delays between the two beams are then created by adding a constant offset $\phi_0$ to every pixel assigned to one of the slopes, for example to $L_1(x)$, $\varphi_1(x) = L_1(x) + \phi_0$
This delay is multiplied in the harmonic generation and can then be measured from the interference in the spectrometer.

% There are, of course, limits to the phase masks that can be created.
% For instance, like many others, our SLMs can only add phases between 0 and \(2\pi\) so the  slopes must be wrapped modulo \(2\pi\).

% Another major limitation is bit-depth.
% Up to this point the masks have been described in a non-discrete manner.
% However, SLMs have limited bit depth which requires the pixel values to be rounded to the nearest bit so they can be displayed on the SLM.

\begin{figure}[ht!]%
\centering
\includegraphics[width=14cm]{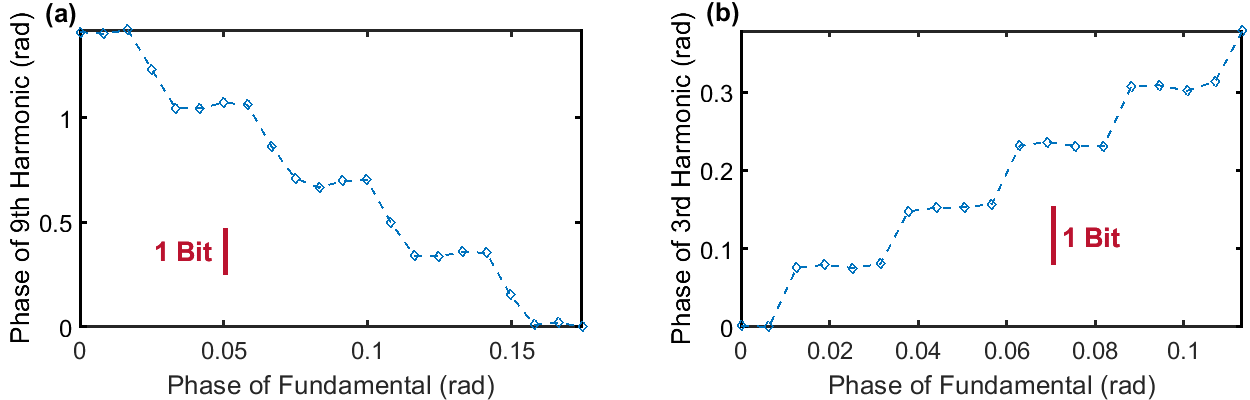}
\caption{
\textbf{(a)} The observed phases of the \(9^{th}\) harmonic generated in ethylene as the phase delay of one beam is scanned.
The red line shows how much this harmonic would be effected by 1 bit of added phase.
There are jumps of one bit twice after every four steps.
This odd jumping pattern happens because the step size chosen hit every 5\(^{th}\) point very well and then round to the other bit for the rest of the cycle.
\textbf{(b)} The observed phase of harmonic 3 generated in quartz as the phase delay of one beam is scanned.
A jump of about 1 bit happens every 4 steps
} \label{steps_fig}
\end{figure}

Although up to this point the masks have been described in a non-discrete manner, the limited bit-depth of SLMs requires the pixel values to be rounded to the nearest bit before they can be displayed on the SLM.
This rounding can eliminate information which leads to a limit for the minimum phase delay, 1 bit.
This is the minimum phase that can be added to each pixel and corresponds to \(\frac{2\pi}{2^8}\) rad = 10.4 as in the case of our 790 nm Meadowlark SLM with a cycle time of 2.6 fs and 18 as for our Hamamatsu SLM at 1400 nm.
Figure \ref{steps_fig} shows a measurement of the 790 nm beam's phase as the phase delay of one beam is scanned.
For these masks the linear slopes were rounded to the nearest bit before having the phase delay added.
The delay was added in steps of 0.0249 rads of the \(9^{th}\) harmonic = 0.34 bits for the Meadowlark and 0.0126 rads of the \(3^{rd}\) harmonic = 0.17 bits for the Hamamatsu.
Clear steps form showing phase jumps of 1 bit for both the Meadowlark (a) and the Hamamatsu (b) SLMs.

% Although 1 bit is the minimum phase that can be added to each pixel, the overall phase of a mask can be manipulated more precisely. 
% \textcolor{red}{add more meat. ie to test our volume averaging scheme or reference intro or ...}

%%%%%%%%%%%%%%%%%%%%%%%%%%%%%%%%%%%%%%%%%%%%%%%%%%%%%%%%%%%%%%%%%%%%%%%%%%%%
\section{Overcoming bit-depth limitations}\label{no_steps}

To create phase effects of less than 1 bit the phases of multiple pixels must be averaged together.
This is accomplished by taking advantage of the rounding that was already being done in the mask making process. 
This also relies on a linear slope always being applied to the SLM to create a zeroth order offset, on top of any other effects desired.
% This is all made possible by the fact that, for ITAS experiments, a linear slope is always applied to the SLM to create a zeroth order offset, on top of any other effects desired.
If a phase effect is added to the mask before it's rounded, it can interact with the fractional part of the slope and cause some pixels to be rounded up instead of down. 
The fractional part of a number x, \(\{ x \}\), is defined as
\begin{equation}
    \{ x \} \equiv x - \mbox{floor}(x).
\end{equation}

The non-fractional part of the phase will add exactly to each pixel so here we'll only consider phase additions of less than 1.
These could be the fractional part of a larger phase effect, separated out to simplify the math.
Assuming the fractional parts of the slope are uniformly and densely distributed in [0,1) then the percent of pixels that flip and round up are equal to the phase added, \(d\).
There are two mechanisms for a phase addition to make a pixel round up.
First if the pixel's original fractional value is below \(\frac{1}{2}\) then \(d\) can increase it above \(\frac{1}{2}\) and cause it to round up to one.
Secondly, if the original value is above \(\frac{1}{2}\) then \(d\) can put it above \(1\frac{1}{2}\) causing it to round to two instead of one.
For a pixel value, \(u\), the chances for its fractional part to be above or below \(\frac{1}{2}\) are 50\%.
If \(\{ u\} < \frac{1}{2}\) and \(d \geq \frac{1}{2}\) then the chances of rounding up are 100\%. 
In this case if \(d < \frac{1}{2}\) then the probability of a round up is
% With this \(\{ u\}\) if \(d < \frac{1}{2}\) then the probability of a round up is
\begin{equation}
    P\left(\{u\} + d \geq \frac{1}{2}\right) = \frac{d}{1/2} = 2d .
\end{equation}
For the second case, where \(\{ u\} \geq \frac{1}{2}\), if \(d < \frac{1}{2}\) then there is no chance of rounding up to two, otherwise the probability is 
\begin{equation}
    P\left(\{u\} + d \geq 1\frac{1}{2}\right) = \frac{d-\frac{1}{2}}{1/2}.
\end{equation}
If we add together the probabilities of the two mechanisms when \(d < \frac{1}{2}\) we get
\begin{equation}
    P(\mbox{pixel increase}) = \left(\frac{1}{2}\right)\cdot(2d) + \left(\frac{1}{2}\right)\cdot(0) = d.
\end{equation}
Adding the probabilities when \(d \geq \frac{1}{2}\) gives 
\begin{equation}
    P(\mbox{pixel increase}) = \left(\frac{1}{2}\right)\cdot(1) + \left(\frac{1}{2}\right)\cdot\left(\frac{d-\frac{1}{2}}{1/2}\right) = d.
\end{equation}
So in the end, no matter the value of \(d\), the probability of a round up increasing the pixel value is \(d\), assuming that the fractional parts of the slope are uniformly distributed in [0,1).
    
\begin{figure}[ht!]%
\centering
\includegraphics[width=\textwidth]{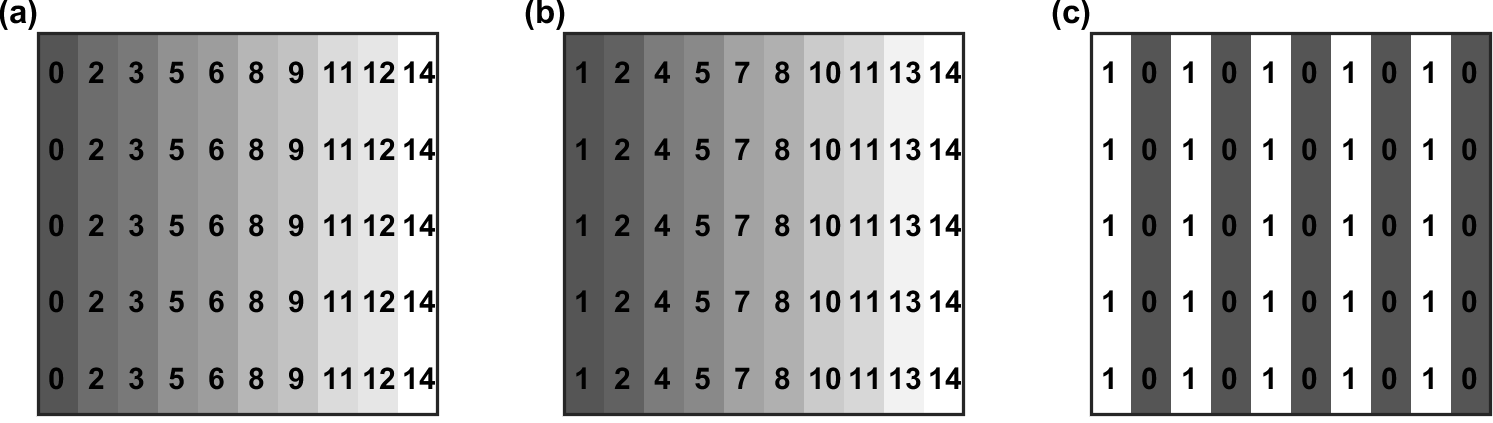}
\caption{
\textbf{(a)} A simple example phase mask. 
It is a single slope angled at  \(1\frac{1}{2}\) bits per pixel.
All values are rounded as they would be in a real use case.
\textbf{(b)} A mask with the same slope but with \(\frac{1}{2}\) bits added before rounding.
\textbf{(c)} The difference between the two masks.
Every other line has its value changed by 1 resulting in a average change of \(\frac{1}{2}\) bits, as desired.
% Here we see that every other line has its value changed by 1 resulting in a average change of \(\frac{1}{2}\) bits, as desired.
} \label{example_fig}
\end{figure}

The assumption of uniformly and densely distributed fractional parts is very commonly true for SLM masks because the slopes will frequently be irrational.
Slopes are commonly defined in radians and then converted to bits bringing in a factor of \(\pi\) with the conversion, making them irrational.
% The slopes will frequently be irrational since they are most commonly defined in radians and the conversion to bits brings in a factor of \(\pi\).
If the slope is irrational then the slopes values will be multiples of an irrational number and multiples of irrational numbers are uniformly distributed and densely packed in [0,1) \cite{Math_book}.
This results in an average phase equal to the desired phase, even if that effect is less than one bit. 
As a simplified example take a mask with a slope of \(1 \frac{1}{2}\) bits per pixel, as shown in Fig.\ref{example_fig}(a), then \(\frac{1}{2}\) a bit can be added to it without changing the slope of the mask, as in Fig.\ref{example_fig}(b).
Figure \ref{example_fig}(c) shows the difference between the two masks; half the pixels have been changed by one resulting in an overall phase delay of \(\frac{1}{2}\), as desired.

% I don't know if this section is necessary anymore or if earlier discussions made it clear that this won't be a problem
This shows that SLMs can add phase delays smaller than 1 bit but doesn't address the smallest delay that is achievable.
There is a limit based on the slope applied, highlighted in Fig.\ref{example_fig} which can apply a phase delay of \(\frac{1}{2}\) bits but could not apply a phase of \(\frac{1}{10}\) bits.
The slope will take values that are multiples of \(1 \frac{1}{2}\) and will only have fractional values of 0 or \(\frac{1}{2}\).
Any addition less than \(\frac{1}{2}\) will have no effect after rounding.
A slope needs to reach fractional values where the phase can change the way the slope is rounded.
% A slope needs to reach fractional values near enough to \(\frac{1}{2}\) for the phase to be able to change the way the slope is rounded.
This limit is, however, reduced significantly for most real-world slopes which, as discussed before, will frequently be irrational and have fractional parts that are uniformly and densely packed in [0,1).
% So the pixels values will generally reach values close to \(\frac{1}{2}\).

There is also a hard lower limit imposed on the averaging effect; at least one pixel must increase its value.
However this limit is 1 bit divided by the number of pixels which is extremely small as most SLMs will have somewhere from tens of thousands to millions of pixels.
% However this limit is extremely small as most SLMs will have somewhere from tens of thousands to millions of pixels and this limit is 1 bit divided by the number of pixels.
The effect is worse for some masks, like those in Fig.\ref{example_fig}, which only have a slope in one direction.
For these masks the columns all have the same values, tying them together.
% This limit can actually be worse for some masks, like those in Fig. \ref{example_fig}.
% This limit should be taken into account though as it is actually worse for some masks, like those in Fig. \ref{example_fig}.
% In these masks there is only a slope in one direction so columns of pixels all have the same values, tying them together.
The limit is then 1 bit divided by the number of different values in the slope.
These masks can be avoided by adding a second slope in the other direction, creating a further zeroth order offset.
% However, this is simple to fix, the masks just need to have a second slope added in the other direction causing a further zeroth order offset.
Now each pixel has a distinct value, although it should be noted that this second slope should depend on another irrational number, not just \(\pi\). 
That guarantees that no pixel values will repeat across the SLM.

A more relevant limit, related to the previous, is the fact that the real-valued phase must be approximated by a fraction of the pixels. 
Small phase effects, closer to the one pixel limit, will be poorly approximated as such.
% This next sentence could use some work (Just take out "over total number of pixels"?)
For arbitrary phases to be hit accurately many pixels must be flipped so that the fraction of pixels flipped over total number of pixels is a good approximation of the phase. 
This is especially important for uses such as ITAS where many phase delay steps will be scanned through in sequence. 
If the desired phase is small, and poorly approximated, each step can vary wildly as it is either rounded up or down to the nearest phase that can actually be displayed by the SLM.
For these purposes it is very important that each step is accurately approximated by a fraction of pixels and the easiest way to achieve this is by having a large number of pixels flip every step.
This error will always be on the order of \(\pm\)one pixel flipping, which as mentioned earlier is a small effect.
\begin{figure}[ht]%
\centering
\includegraphics[width=\textwidth]{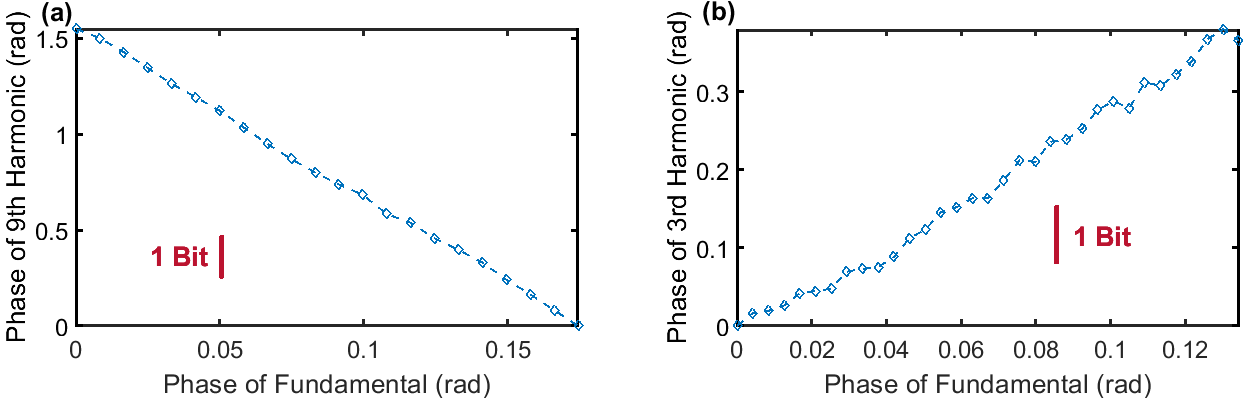}
\caption{
\textbf{(a)} The observed phases of harmonic 9 generated in Ethylene as the phase delay of one beam is scanned using the Meadowlark SLM.
They are now smooth lines with a steps size of 0.0249 rads of the 9\(^{\mbox{th}}\) harmonic = 0.34 bits.
The red line shows what phase, in radians of the harmonic, the harmonic would gain from a 1 bit addition on the SLM.
\textbf{(b)} The observed phase of harmonic 3 generated in quartz as the phase delay of one beam is scanned using the Hamamatsu SLM.
This is a smooth line with a step size of 0.0126 rads of the 3\(^{\mbox{rd}}\) harmonic = 0.17 bits.
} \label{smooth_fig}
\end{figure}
%Double check these slopes
To get more specific, masks were generated with a slope of \(0.124\) rads per pixel (approximately \(20\pi\) in total) in one direction and \(\sqrt{2}/300\) rads per pixel in the other which are characteristic numbers for our ITAS setup. 
% To get more specific we generated masks with a slope of \(0.124\) rads per pixel (approximately \(20\pi\) in total) in one direction and \(\sqrt{2}/300\) rads per pixel in the other which are characteristic numbers for our ITAS setup. 
For a typical 8-bit SLM with 500x500 pixels we conservatively estimate that light is only incident on 300x300 of them, resulting in a minimum step size of 0.11 zs for 800 nm light if one pixel changes.
If 500 of these masks are generated in sequence each with an additional phase delay of \(0.001\mbox{ bits} = 10\mbox{ zs}\), which is approximately 100 pixels flipping, then the masks have an average delay step of \(10\mbox{ zs} \pm 0.27\mbox{ zs}\), an error of about 2.5 pixels. % 0.001 bits \pm 0.000027 bits
% If we generate 500 of these masks in sequence each with an additional phase delay of \(0.001\mbox{ bits} = 10\mbox{ zs}\), which means about 100 pixels flipping, then we see the masks have average delays of \(10\mbox{ zs} \pm 0.27\mbox{ zs}\).% 0.001 bits \pm 0.000027 bits
While the exact error of these phases varies based on what specific phase delay is desired and what slopes are applied the error is always similar and on the order of a few pixels flipping incorrectly.
% While the exact error of these delays varies based on what specific delay is desired and what slopes are applied the error is always similar and on the order of a few times the 1 pixel minimum.
This SLM could reliably take steps of one thousandth of a bit, making its effective bit-depth 18, far exceeding its expected performance.
For a higher spatial resolution SLM (more pixels) these gains are even more impressive; the Meadowlark SLM with 1920x1152 pixels can be reduced, without too much spatial resolution loss, to 800x800 pixels.
An 8-bit, 800x800 SLM has a minimum step size of 0.015 zs for 800 nm light if one pixel flips.
This gives another order of magnitude gain, meaning steps of 1 zs could be reliably taken, which is far beyond what we have the ability to measure.

Experimental demonstrations of the averaging effect were tested with two of our SLMs.
% To justify the averaging effect this technique uses it was tested with two of our SLMs.
In Fig.\ref{smooth_fig} we show phase scans with step sizes smaller than 1 bit with both the Meadowlark (a) and Hamamatsu (b) SLMs.
The Meadowlark used a step size of 0.34 bits = 0.0083 rads of the fundamental while the Hamamatsu had steps of 0.17 bits = 0.0042 rads of the fundamental.
Phase control exceeding 1 bit is clearly demonstrated. 
% For the Meadowlark we used a step size of 0.34 bits = 0.0083 rads of the fundamental and with the Hamamatsu steps of 0.17 bits = 0.0042 rads of the fundamental.
These step sizes were chosen as they are close to the limits of what can be measured in these setups due to noise sources such as vibrations and beam pointing.
% The Meadowlark's scan deviated from the expected linear slope with an RMS of 0.06 bits which is significantly larger than 
Our experimental results show that, at least for linear functions, the required averaging does happen and details of less than 1 bit can be applied to phase masks with measured optical effects.
% Although this was only tested with ITAS setups the technique used should be widely applicable as it relies only on having a zeroth order offset.

% \textcolor{red}{Make note of Tobi's zoned image idea to make any image better. trading off res for bitdepth. here or in next section?}
\begin{figure}[ht]%
\centering
\includegraphics[width=\textwidth]{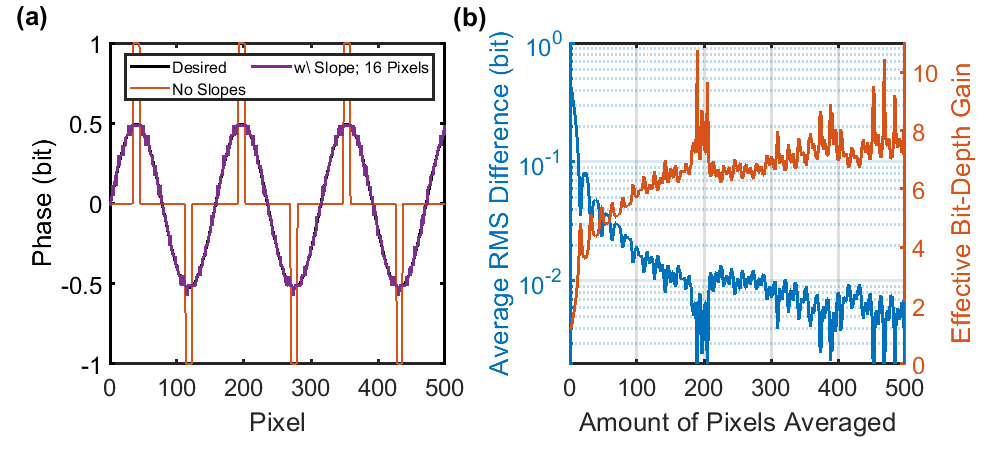}
\caption{
\textbf{(a)} 
A simulated sine wave phase effect of amplitude 0.51 bits applied along one axis of a 500x500pixel SLM.
In black is the desired phase that would be applied to each row of the SLM.
Red shows what phase would be applied without this technique. 
Only the peaks and valleys of the wave are large enough to be displayed. 
% The purple line uses this technique and shows what phase is displayed if the linear slopes are subtracted off and the center 16 rows of the mask are averaged together.
The purple line shows what phase is displayed if the linear slopes are subtracted off and the center 16 rows of the mask are averaged together.
The slopes used were \(\pi\slash50\) and \(\sqrt{2}\slash50\) bits per pixel.
There is a significant increase to the precision of phase that can be applied when pixels are averaged together.
\textbf{(b)}
How the precision of the applied phase increases when more pixels are averaged together. 
Blue shows the average RMS difference between the desired phase and the averaged phase. 
Red shows the equivalent bit-depth increase that would be needed to apple the phase this precisely.
} \label{structures_fig}
\end{figure}
Although this method was only experimentally tested with flat phase offsets it can be applied to more complicated phases. 
Indeed, any phase function can always be approximated to discrete sections of pixels that are relatively flat in phase.
Depending on the specific effect in question these sections might be small, only a few pixels, or much larger, with hundreds of pixels depending on the ideal compromise between spatial and phase resolution.
The gain in phase resolution scales roughly with $N_p$, the number of pixels used in the average.
% In either case, the phase of those sections should be more precisely displayed on the SLM when linear slopes are applied for all the reasons listed above.
When all sections are combined together the total phase effect should be more precisely represented on the SLM, though this improvement is difficult to quantify as it depends heavily on the specific effect being applied.
This enhancement is already helping any system that uses linear slopes for a zeroth order offset but ensuring that slopes are applied along both SLM axes and that the slopes depend on distinct irrational numbers is a simple change that will increase these benefits.

To test this technique with more complicated phase masks we simulated a mask with a sine wave of amplitude 0.51 bits along one axis of a 500x500 SLM. The desired phase function is shown in black in Fig. \ref{structures_fig} (a).
As shown in Fig. \ref{structures_fig}(a) in red, an 8-bit representation poorly approximates the true sine wave.
When irrational slopes, of \(\pi\slash50\) and \(\sqrt{2}\slash50\) bits per pixel, are applied and 16 pixels along the other axis are averaged together the wave is well approximated, shown in purple. 
% This will also work for any desired phase effect that can be divided into sections of 16 pixels of equal value.
% Fig. \ref{structures_fig}(b) shows how the average number of bits between the displayed phase and desired phase decreases as more pixels are averaged.
Fig. \ref{structures_fig}(b) shows, in blue, how the average root mean squared (RMS) difference between the displayed phase and desired phase decreases as more pixels are averaged together.
Also shown, in red, is the equivalent increase in bit-depth that would be required to display the mask that precisely.
For full 2-D phases instead of pixels in a line arbitrary 2D sections of pixels with similar values have to be used.

%%%%%%%%%%%%%%%%%%%%%%%%%%%%%%%%%%%%%%%%%%%%%%%%%%%%%%%%%%%%%%%%%%%%%%%%%%%%
\section{Conclusion}

% We've presented a technique that allows SLMs to be used with higher precision than their bit-depth would seem to allow and tested it with multiple SLMs when using them to generate harmonics in ITAS measurements.
We have presented a technique that allows SLMs to overcome their bit-depth limitations and experimentally tested it with multiple SLMs.
% All that is required for this method is a system that uses the average of the phase and not the per pixel values. % delay as well as a zeroth order offset with an irrational factor applied to the SLM.
% This was tested with multiple SLMs from two major manufacturers and shown to be effective when using SLMs to generate harmonics as in ITAS measurements.
The method relies on using zeroth order offsets yielding irrationally valued slopes and mask values that are uniformly distributed and with densely packed fractional parts when averaged locally over a subset of pixels.
The fractional parts of any additional phase, when locally averaged with such slopes, are experimentally measured to have features of higher precisions than the bit-depth. 
This was examined in detail in attosecond interferometric measurements where the minimum phase delay was improved by a factor of 3, and theoretically predicted to allow an 8-bit SLM to become effectively 18-bit.
With this method two limiting factors of SLMs are accounted for at once; elimination of the zeroth order reflection and overcoming bit-depth limits. Increase in phase resolution comes at the cost of spatial resolution (less effective pixels) but does not affect the bit rate and the refresh speed. Our approach matches modern SLM trends of increased number of pixels, faster bit rates but decreased bit depth and thus can lead to a general approach in the industry.

%%%%%%%%%%%%%%%%%%%%%%%%%%%%%%%%%%%%%%%%%%%%%%%%%%%%%%%%%%%%%%%%%%%%%%%%%%%%

\section{Acknowledgements}
We thank fruitful discussions with Peter Brunt of AVR Optics. Research supported by US Department of Energy,
Office of Science, Chemical Sciences, Geosciences, \& Biosciences Division grant DE-SC0019098. B. D. was supported by the Office of Naval Research, Directed Energy Ultra-Short Pulse Laser Division grant N00014-19-1-2339. 

%%%%%%%%%%%%%%%%%%%%%%%%%%%%%%%%%%%%%%%%%%%%%%%%%%%%%%%%%%%%%%%%%%%%%%%%%%%%

\section{Disclosures}
The authors declare no conflicts of interest.

%%%%%%%%%%%%%%%%%%%%%%%%%%%%%%%%%%%%%%%%%%%%%%%%%%%%%%%%%%%%%%%%%%%%%%%%%%%%

\par\medskip\noindent{\bfseries Data availability. \enspace}
Data underlying the results presented in this paper are not publicly available at this time but may be obtained from the authors upon reasonable request.

%%%%%%%%%%%%%%%%%%%%%%%%%%%%%%%%%%%%%%%%%%%%%%%%%%%%%%%%%%%%%%%%%%%%%%%%%%%%

\bibliography{bibly}

\end{document}